\documentclass[prl,twocolumn,showpacs,superscriptaddress,a4paper]{revtex4} 
\usepackage{amsfonts,amsthm,amssymb,latexsym,graphicx,color,layout,psfrag}
\usepackage[nointlimits]{amsmath}

\newcommand{\mc}[1]{{\mathcal #1}}
\renewcommand{\epsilon}{\varepsilon}

\begin{document}

%
\title{On the long range correlations of thermodynamic systems out of
  equilibrium}   

\author{L. Bertini}
\affiliation{Dipartimento di Matematica, Universit\`a di Roma
``La Sapienza", P.le A. Moro 2, 00185 Roma, Italy}

\author{A. De Sole}
\affiliation{Dipartimento di Matematica, Universit\`a di Roma
``La Sapienza", P.le A. Moro 2, 00185 Roma, Italy \\
and Department of Mathematics, Harvard University, 
1 Oxford St., Cambridge, MA 02138, USA
}

\author{D. Gabrielli}
\affiliation{Dipartimento di Matematica, Universit\`a dell'Aquila,
67100 Loc. Coppito, L'Aquila, Italy}

\author{G. Jona-Lasinio}
\affiliation{Dipartimento di Fisica and INFN,
Universit\`a di Roma ``La Sapienza", P.le A. Moro 2, 00185 Roma,
Italy}

\author{C. Landim}
\affiliation{IMPA, Estrada Dona Castorina 110, J. Botanico, 22460 Rio
de Janeiro, Brazil, \\ and CNRS UMR 6085, Universit\'e de Rouen, 
F76801 Saint-\'Etienne-du-Rouvray, France}


%
%

\begin{abstract}
Experiments show that macroscopic systems in a stationary
nonequilibrium state exhibit long range correlations of the local
thermodynamic variables.  In previous papers we proposed a
Hamilton-Jacobi equation for the nonequilibrium free energy as a basic
principle of nonequilibrium thermodynamics.  We show here how an
equation for the two point correlations can be derived from the
Hamilton-Jacobi equation for arbitrary transport coefficients for
dynamics with both external fields and boundary reservoirs.  In
contrast with fluctuating hydrodynamics, this approach can be
used to derive equations for correlations of any order.
Generically, the solutions of the equation for the correlation
functions are non-trivial and show that long range correlations are
indeed a common feature of nonequilibrium systems. Finally, we
establish a criterion to determine whether the local thermodynamic
variables are positively or negatively correlated in terms of
properties of the transport coefficients.
\end{abstract}
%
%


\maketitle

%
%

The basic idea in the construction of nonequilibrium thermodynamics is
that of local equilibrium.  This means that on the macroscopic scale
it is possible to define local thermodynamic variables like density,
temperature, or chemical potentials, which vary smoothly.
Microscopically this implies that the system reaches local equilibrium
in a time which is short compared to the times typical of macroscopic
evolution as described, for example, by hydrodynamic equations.  Even
if the local behavior of a nonequilibrium system is well described by
a Gibbs state, if we probe the system at distances comparable to its
macroscopic size, relevant differences with respect to equilibrium do
appear.  It is indeed an established fact that macroscopic systems in
a stationary nonequilibrium state generically exhibit long range
correlations of the local thermodynamic variables.  For a review of
experimental results see \cite{DKS}. The theories of correlations in
thermodynamic systems out of equilibrium that can be found in the
literature are mainly based on the so called fluctuating hydrodynamics
\cite{LL,S1,SC,K}, that is on a linearization around a steady state of
the hydrodynamic equations perturbed by random currents. For diluted
systems this problem has also been approached within kinetic theory
\cite{CDK}.  

It has been shown for several stochastic dynamics that the
nonequilibrium free energy, which is the generating functional of the
correlation functions, satisfies a Hamilton-Jacobi equation
\cite{BDGJL1, BDGJL2}.  The basic feature of this equation is that it
does not depend on the details of the microscopic dynamics but only on
the macroscopic transport coefficients, i.e. the diffusion coefficient
and the mobility. It can thus be considered as a general principle of
nonequilibrium thermodynamics.  In this paper we show how the
Hamilton-Jacobi equation can be used to study correlation functions.
In contrast with fluctuating hydrodynamics, this approach can be used
to derive equations for correlations of any order.  For simplicity, we
consider diffusive systems in arbitrary dimension described by a
single thermodynamical variable with both external fields and boundary
reservoirs.  We show that the two point function satisfies a linear
partial differential equation which depends, as in fluctuating
hydrodynamic, on the transport coefficients evaluated in the
stationary state.  These coefficients can be computed from the
microscopic dynamics or can be obtained from experiments.
Generically, the solutions of the equation for the correlation
functions are non-trivial and show that long range correlations are a
common feature of nonequilibrium systems. We remark that the long
range correlations arise from nonlinear effects in the hydrodynamic
equation.  One of the most interesting consequences of the equation
for the correlations is that it is possible to establish whether the
fluctuations are positively or negatively correlated in terms of
structural properties of the transport coefficients.

Systems for which the above scheme can be applied are characterized by
a separation of scales both in space and time.  We analyze these
systems in the limit in which this separation becomes sharp.  In the
following we let $\epsilon$ be the ratio between microscopic scale,
say of order of the intermolecular distance, and the linear size of
the system $L_0$.  Let us consider a macroscopic system in
a domain $\Lambda$ in space dimension $d$ described in terms of a
single thermodynamic observable $\rho$, that we think of as the
density of particles.  We denote by $\rho(x)$ the local density at the
macroscopic point $x$, obtained as the density of particles in a
volume of side length $\ell$, with $\epsilon L_0 \ll \ell \ll L_0$.
We then introduce the corresponding ensemble $P_\epsilon$ as the
distribution of such local density. This distribution is inherited
from the microscopic ensemble. 

According to the Boltzmann--Einstein theory of equilibrium
thermodynamic fluctuations \cite{E} the relationship between the
free energy $F$ and the probability distribution $P_\epsilon$ is the
following
\begin{equation}
\label{fe}
P_\epsilon(\rho(x),\,x\in \Lambda) \propto e^{-F(\rho)/\epsilon^d}
\end{equation}
where we absorbed the temperature $T$, fixed in the one phase region, 
in the definition of $F$.  
In this equilibrium setting the free energy $F$ is a local functional,
i.e.\ it can be written as
\begin{equation}
\label{feeq}
F(\rho) \,=\, \int_\Lambda\!dx \,  f(\rho(x)) 
\end{equation}
where $f(\rho)$ is the free energy per unit volume.
In particular, there are no correlations, 
i.e fluctuations of $\rho(x)$ and $\rho(y)$, for $x\neq y$, are independent.

For systems out of equilibrium, e.g.\ when they are in contact with
reservoirs and/or under the action of an external field $E$, we let
$P_\epsilon$ be the corresponding stationary ensemble. For such
systems we define the associated nonequilibrium thermodynamic
functional $F(\rho)$ by \eqref{fe}.  That this is the appropriate
definition is supported by the following facts discussed in
\cite{BDGJL1, BDGJL2}. First, the stationary state corresponds to
the absolute minimum of $F$, thus generalizing the maximum entropy
principle of equilibrium statistical mechanics. Moreover, the
derivative of $F$ with respect to $\rho$ provides the correct
definition of thermodynamic force, that is the force responsible for
dissipation, as shown in \cite{BDGJL2}.

In order to present the Hamilton-Jacobi equation for the free energy,
we introduce the dynamical behavior of the system. The macroscopic
evolution of the density is described by a (in general nonlinear)
diffusion equation with a transport term corresponding to the external
field $E$, namely
\begin{equation}
\label{hyde}
\partial_t \rho(t,x) = 
  \partial_{i} \Big[ \, \frac 12 \,D_{ij}(\rho)
  \partial_{j} \rho(t,x) -\chi_{ij}(\rho) E_j(x) \Big]
\end{equation}
where we use Einstein's convention of summing over repeated indices
and $\partial_i$ stands for $\partial/\partial x_i$, $i=1,\cdots,d$.
In \eqref{hyde} $D$ is the diffusion coefficient and $\chi$ is the
mobility.  This equation has to be supplemented by the appropriate
boundary conditions.  For nonequilibrium systems in contact with
particle reservoirs, this amounts to fix the value of $\rho$ at the
boundary of $\Lambda$.  We denote by $\bar\rho(x)$ the stationary
density profile, i.e. the unique stationary solution of \eqref{hyde}.
This is the density profile in which the nonequilibrium free energy
$F$ attains its minimum.  Equation \eqref{hyde} identifies the
current $\bar J$ of particles flowing though the system in the
stationary profile as
\begin{equation}
\label{current}
{\bar J}_i(x)\,=\,-\Big[ \,\frac 12 \,D_{ij}(\bar\rho) \partial_{j}
\bar\rho(x)   -\chi_{ij}(\bar\rho) E_j(x) \Big]
\end{equation}
The equation for $\bar\rho$ then reads $\partial_{i} \bar J_i(x)=0$,
i.e.\ $\bar J$ is divergenceless.

As shown in \cite{BDGJL1,BDGJL2}, the nonequilibrium free energy $F$
is the maximal solution of the Hamilton-Jacobi equation
\begin{equation}
\label{HJnoE}
\Big\langle \partial_i \frac {\delta F}{\delta \rho} \,,\, 
\frac 12 \chi_{ij} (\rho) \partial_j \frac {\delta F}{\delta \rho} 
-  \frac 12 D_{ij}(\rho)\partial_j \rho  +  \chi_{ij}(\rho) E_j 
\Big\rangle = 0
\end{equation} 
where $\langle f,g\rangle = \int_{\Lambda}\! dx \, f(x) g(x)$ and $F$
satisfies the boundary condition $\delta F / \delta \rho
\big|_{\partial\Lambda} =0$.  The derivation of this equation is based
on a dynamical argument. One first generalizes the Boltzmann-Einstein
formula \eqref{fe} to space-time trajectories obtaining the asymptotic
probability of deviation from solutions of the hydrodynamic equation
\eqref{hyde}. The functional $F$ can be identified as follows: if at
$t=-\infty$ the system is in the stationary profile $\bar\rho$ then
the probability of observing the profile $\rho$ at $t=0$ is
proportional to $\exp\{ - \epsilon^{-d} F(\rho)\}$. The free energy
$F$ solves a variational problem whose associated Hamilton-Jacobi
equation is \eqref{HJnoE}.  The validity of \eqref{HJnoE} has been
established for stochastic lattice gases for which the local
equilibrium can be proven.

In the case of equilibrium systems, in which $E=0$ and
$\rho|_{\partial\Lambda}$ is constant, the profile $\bar\rho$ is
constant; it is then simple to check that the solution of
\eqref{HJnoE} has the form \eqref{feeq} where the free energy per unit
volume $f(\rho)$ satisfies the Einstein relation $D_{ij}(\rho) =
\chi_{ij}(\rho) f''(\rho)$, normalized so that
$f'(\bar\rho)=f(\bar\rho)=0$.  For nonequilibrium systems the
solutions of \eqref{HJnoE} cannot, in general, be obtained in a closed
form.  For special choices of the transport coefficients, which
correspond to well studied stochastic dynamics, the solution of
\eqref{HJnoE} can be expressed in terms a non linear boundary value
problem, see \cite{BDGJL2,DLS2,BGL,DE}.  Contrary to equilibrium, a
general feature of the nonequilibrium free energy $F$ is that it is
not a local functional.  This implies that macroscopic density
correlations do generically appear.

The density-density correlation is defined as
\begin{equation}
\label{C}
C(x,y) \,\approx\,  
\epsilon^{-d} \int  \big(\rho(x)-\bar\rho(x)\big)\big(\rho(y)-\bar\rho(y)\big) dP_\epsilon
\end{equation}
Note that the scaling $\epsilon^{-d}$ corresponds to the Gaussian
fluctuations.  This function is related to the non
equilibrium free energy $F$ by
\begin{equation}
\label{c-1}
C^{-1}(x,y) \,=\, \frac{\delta^2F(\rho)}{\delta\rho(x)\delta\rho(y)}
\Big|_{\rho=\bar\rho} 
\end{equation}
so that $F(\rho) = 
(1/2) \, \langle (\rho-\bar\rho), C^{-1}(\rho-\bar\rho) \rangle 
+ o((\rho-\bar\rho)^2)$. 
We next introduce the function $B(x,y)$ by
\begin{equation}\label{alb4}
C(x,y) = C_{eq}(\bar\rho(x))\delta(x-y) + B(x,y)\,\,,\,\,\,\, x,y\in\Lambda
\end{equation}
where, as $f$ is the equilibrium free energy per unit volume,  
$C_{eq}(\rho)=f''(\rho)^{-1}$ gives the (local) equilibrium variance. 
Notice that, while $D$ and $\chi$ are $d\times d$
matrices, $C_{eq}$, $C$ and $B$ are multiples of the
identity.  Since the density fluctuations at the boundary are
determined only by the reservoirs, we have that $B(x,y) =0$ when
either $x$ or $y$ is at the boundary of $\Lambda$.

In order to write an equation for $B$, introduce the elliptic
differential operator $L$ as
\begin{equation}
\label{L}
L\,=\,
\frac12 \, D_{ij}(\bar\rho(x)) \, \partial_{i} \partial_{j} +
\chi^\prime_{ij}(\bar\rho(x)) E_j(x) \, \partial_{i} 
\end{equation}
and $\mc L=L\oplus L$, i.e.\ $\mc L$ satisfies $\mc
L\varphi(x)\psi(y)=\psi(y) L\varphi (x) + \varphi(x) L\psi(y)$.  Let
finally $L^\dagger$ be the adjoint of $L$ with Dirichlet boundary
conditions on $\Lambda$, i.e.
\begin{eqnarray*}
L^\dagger\varphi(x) &=&
\frac12 \partial_{i} \partial_{j}
\big[ D_{ij}(\bar\rho(x))  \varphi(x)\big] \\
&&\,\,\,-\, \partial_{i}\big[ 
\chi^\prime_{ij}(\bar\rho(x))E_j(x) \varphi(x)\big]
\end{eqnarray*}
and $\mc L^\dagger=L^\dagger\oplus L^\dagger$.
As shown below, from the Hamilton-Jacobi equation \eqref{HJnoE} it
follows that $B$ satisfies 
\begin{equation}
\label{alb7}
\mc L^\dagger B(x,y) = - h(x) \delta(x-y)
\end{equation} 
where
\begin{equation}
\label{alb6.5} 
h(x) = - \partial_{i} 
\big[ \chi^\prime_{ij}(\bar\rho(x)) D^{-1}_{jk}(\bar\rho(x)) \, {\bar
    J}_k(x) \big] 
\end{equation} 
The choice of the  Dirichlet boundary condition for $L$ 
corresponds to the vanishing of $B$ at the boundary.

We draw some important consequences of equation \eqref{alb7}.  Since
$D$ is positive definite, $\mc L$ is an elliptic operator in
$\Lambda\times\Lambda$ with Dirichlet boundary condition.  Let
$G(x,y;,x',y')$ be its Green function, i.e.\ the solution of $-\, \mc
L \, G(x,y;x',y') = \delta(x-x')\delta(y-y')$.  Then the solution of
\eqref{alb7} is
\begin{equation}\label{sol}
B(x,y)  = \int_{\Lambda} \! dz \, G(z,z;x,y)  \, h(z)
\end{equation}
Since $G\ge 0$, we conclude that if $h\ge 0$, respectively $h\le 0$,
fluctuations of the density are positively correlated, i.e.\ $B\ge 0$,
respectively negatively correlated, i.e.\ $B\le0$.  In
absence of external field, i.e.\ for $E=0$, we have, from
\eqref{current} and \eqref{alb6.5}, that $h(x)= \frac12\partial_{i}
\partial_{j} \chi_{ij} (\bar\rho(x))$.  If we further assume that
$\chi$ is a multiple of the identity, we get that $B \ge 0$,
respectively $B \le 0$, if $\chi(\bar\rho(x))$ is a subharmonic,
respectively superharmonic, function of $x$, namely $\Delta
\chi(\bar\rho(x))$ is positive, respectively negative.

For equilibrium systems the current $\bar J$ in the stationary profile
vanishes, hence $h=0$ so that $C=C_{{eq}}$.  This conclusion is also
true for systems with periodic boundary conditions and constant
external field.  In such case $\bar\rho$ and $\bar J$ are constant so
that $h$ vanishes.  More generally, in \cite{NOI9} it is shown that,
in the periodic case, the nonequilibrium free energy $F$ is the same
as the equilibrium one. For isotropic systems $D$ and $\chi$ are
multiples of the identity, in such a case $h$ vanishes if and only if 
$\partial_i \big[ \chi'(\bar\rho(x)) D^{-1}(\bar\rho(x)) \big] \,
\bar J_i(x)=0$  for any $x$ in $\Lambda$. 
For instance, in the zero range process $\chi$ is an increasing 
function and $D=\chi'$, so that $h$ vanishes and the
model does not exhibit long range correlations.

We now consider some one-dimensional systems and choose
$\Lambda=(0,1)$.  We let $\rho(0)=\rho_0\le\rho_1=\rho(1)$ be the
boundary conditions imposed by the reservoirs.  The one-dimensional
exclusion process has $0\le\rho\le 1$, $D(\rho)=1$,
$\chi(\rho)=\rho(1-\rho)$.  For $E=0$ equation \eqref{alb7} implies
that $B$ is proportional to the Green function of the Dirichlet
Laplacian in $(0,1)$, namely
\begin{equation}
\label{covsep}
B(x,y)\,=\,-(\rho_1-\rho_0)^2 \,  x(1-y)\,,\qquad  0\le x\le y\le 1
\end{equation}
This result was first derived in \cite{S1}.  Notice that, in agreement 
with the above discussion, since $\chi$ is concave and $\bar\rho'$ is
constant, $B\le0$.  For constant external field $E$ the solution of
\eqref{alb7} is given by
\begin{equation}
\label{covsepE}
B(x,y)=2\bar J \, \bar\rho'(x)\bar\rho'(y)
\frac{\int_0^x\!du\,\bar\rho'(u)^{-1}
\int_y^1\!du\,\bar\rho'(u)^{-1}}{\int_0^1\!du\,\bar\rho'(u)^{-1}}  
\end{equation}
for $0\le x\le y\le 1$.  This formula for the correlation function has 
been derived in \cite{DELO}. It can be shown that $\bar\rho$ is
increasing.  Therefore the correlations $B(x,y)$ have the same sign as
$\bar J$.  We finally discuss the correlations in the
Kipnis-Marchioro-Presutti model \cite{KMP,BGL}. This model describes a
chain of one-dimensional harmonic oscillators which are mechanically
uncoupled and interact by exchanging stochastically the energy with
the neighboring sites. Accordingly, the thermodynamic variable $\rho$
is the energy density. For this model $D(\rho)=1$ and
$\chi(\rho)=\rho^2$. The solution of \eqref{alb7} is then the same as
the one for the exclusion model, i.e.\ \eqref{covsep}, but with the
opposite sign. For this model the energy fluctuations are therefore
positively correlated, in agreement with the above discussion and the
fact that $\chi$ is a convex function. Since $B$ is proportional to
Green function of the Dirichlet Laplacian, we also have that it is a
positive definite operator. This means that, in the quadratic
approximation near $\bar\rho$, the nonequilibrium free energy is
smaller than the local equilibrium one.  The opposite behavior takes
place in the exclusion process. For these models this result has been
proven also for large fluctuations, that is beyond the quadratic
approximation \cite{BDGJL2,BGL,DLS2}.

We here show that the above results can be seen as special cases of a
class of one-dimensional systems for which \eqref{alb7} can be solved
explicitly.  Assume that 
\begin{equation}
\label{unphyscond}
2\chi'(\bar\rho(x) )  E(x) h(x) = \partial_x [ D(\bar\rho(x)) h(x) ] 
\end{equation}
holds for any $x$ in $(0,1)$. Then it is simple to check that the operator  
$\hat L = -2 h(x)^{-1} L^\dagger$ is self-adjoint and 
$B$ is its Green function.  
By standard Sturm Liouville theory we then get 
\begin{equation}
\label{sturm}
B(x,y) = \frac 1K \, \alpha_1 (x) \alpha_2 (y), \qquad  0\le x\le y \le 1
\end{equation}
where $\alpha_i$, $i=1,2$, solves the Cauchy problem $\hat L \alpha_i
=0$, with $\alpha_1(0)=0$, $\alpha'_1(0)=1$ and $\alpha_2(1)=0$,
$\alpha'_2(1)=1$.  Finally $K= - [ D(\bar\rho(x))/h(x) ]
W(\alpha_1,\alpha_2)$, where $W$ is the Wronskian, is constant in
$x$. Both for the exclusion and the KMP processes \eqref{unphyscond}
holds for any constant external field $E$. Simple computations then
yield \eqref{covsep} and \eqref{covsepE} as special cases of
\eqref{sturm}.  Moreover, from \eqref{sturm} we also get the
correlation function $B(x,y)$ for the KMP model with constant external
field, which is the same as \eqref{covsepE} with opposite sign.

To conclude we show how equation \eqref{alb7} can be derived from the
Hamilton-Jacobi equation for the nonequilibrium free energy.
To derive the equation for the correlations, 
it is convenient to introduce $G(\lambda)$ as the Legendre transform of
$F(\rho)$, i.e.\ $G(\lambda) = \sup_\rho \big\{\langle
\lambda,\rho\rangle - F(\rho) \big\}$. The function $\lambda(x)$, $x$ in 
$\Lambda$, can be interpreted as the variation of the chemical
potential, in particular $\lambda=0$ corresponds to the stationary profile
$\bar\rho$. 
By Legendre duality, equation \eqref{HJnoE} is then equivalent to 
\begin{eqnarray*}
\!\!\!\!\!\!\!\!\!\!\!\!\!\!\!\!\! &&
\frac 12  \Big\langle \partial_i \lambda\, , \,  
\chi_{ij} \Big(\frac {\delta G}{\delta \lambda}\Big) 
\partial_j \lambda \Big\rangle \\
\!\!\!\!\!\!\!\!\!\!\!\!\!\!\!\!\! && \quad
-\; \Big\langle \partial_i \lambda\, , \, 
\frac 12 D_{ij}\Big(\frac {\delta G}{\delta \lambda}\Big) 
\partial_j \frac {\delta G}{\delta \lambda}  - 
\chi_{ij}\Big(\frac {\delta G}{\delta \lambda} \Big) E_j 
\Big\rangle = 0
\end{eqnarray*}
for any $\lambda$ which vanishes at the boundary of $\Lambda$. 
In the quadratic approximation we have 
$G(\lambda) =  \langle \lambda,\bar\rho \rangle 
+  (1/2)\,  \langle \lambda,C \lambda\rangle +o(\lambda^2)$.
Since $\bar\rho$
is the stationary solution of \eqref{hyde},
the first order term in $\lambda$ vanishes in the above equation,
so that $C$ solves 
\begin{equation*}
\Big\langle \partial_i \lambda \, ,\,  
 \frac 12 \chi_{ij} (\bar\rho) \partial_j \lambda - 
\frac 12 \partial_j \big( D_{ij}(\bar\rho) C \lambda \big) 
+  \chi'_{ij}( \bar\rho) E_j C \lambda  \Big\rangle = 0
\end{equation*}
By plugging \eqref{alb4} in the above equation, recalling \eqref{current} and 
$D^{-1}_{ik}(\bar\rho) \chi_{kj}(\bar\rho) = C_{eq}(\bar\rho)
\delta_{ij}$, we get 
\begin{eqnarray*}
&& \Big\langle \partial_i \lambda \, ,\,  
- \frac 12 \partial_j \big( D_{ij}(\bar\rho) B \lambda \big) 
+  \chi'_{ij}( \bar\rho) E_j B \lambda  
\Big\rangle 
\\
&&\qquad
= -\Big\langle \partial_i \lambda \, ,\,  
\chi'_{ij}( \bar\rho) D^{-1}_{jk} (\bar\rho) \bar J_k \lambda  
\Big\rangle\qquad\qquad \qquad 
\end{eqnarray*}
Integrating by parts we then get 
\begin{eqnarray*}
&&\Big\langle \lambda \, ,\,  
 \frac 12 \partial_i \partial_j 
\big( D_{ij}(\bar\rho) B \lambda \big) -  
\partial_i \big( \chi_{ij}' (\bar\rho) E_j  B \lambda \big)
\Big\rangle 
\\
&& \qquad  
= \frac 12 
\Big\langle \lambda \, ,\, 
\Big[ \partial_i \big( \chi'_{ij}(\bar\rho) D^{-1}_{jk} 
\bar J_k \big) \Big] \lambda \Big\rangle 
 \qquad \qquad \qquad 
\end{eqnarray*}
Recalling \eqref{L} and \eqref{alb6.5}, 
the above equation can be written in an operator notation as  
\begin{equation*}
 L^\dagger B  + B L  = - h  
\end{equation*}
Here $h$ denotes a multiplication operator and we used that $B(x,y)=B(y,x)$. 
Denote by $L_x^\dagger$, respectively $L_y^\dagger$, the operator $L^\dagger$
acting on functions of $x$, respectively of $y$. From the previous equation
we get that for any function $\varphi$ vanishing on $\partial\Lambda$ we have 
\begin{equation*}
\int_{\Lambda}\!dy \, \big[ 
L^\dagger_x B(x,y) \varphi(y) + L^\dagger_y B(x,y) \varphi(y) 
\big] =  - h(x) \varphi(x)
\end{equation*}
that is \eqref{alb7}.

We conclude with some comments on the applicability of equation (10).
The Hamilton-Jacobi equation from which it is derived was originally
introduced in the study of microscopic models whose macroscopic
behavior is of diffusive type and there are no normal modes like sound
waves.  This case covers for example nonequilibrium solutions of
different chemical species but no chemical reactions.  The
Hamilton-Jacobi equation is strictly connected with the dynamic
generalization of the Einstein formula for static fluctuations
obtained in \cite{BDGJL1,BDGJL2} and it has been extended to
systems without conservation laws \cite{NOI9, BJ} where the
hydrodynamic equations are of reaction-diffusion type, as it happens
in presence of chemical reactions.  It can be generalized to
situations with several conservation laws where one of the
hydrodynamic equations is Navier-Stokes.  We finally emphasize that
the approach here presented, based on an exact equation for the
nonequilibrium free energy which takes into account the nonlinear
effects in the systems, can be applied to derive equations for higher
order correlations.


\begin{thebibliography}{99}


\bibitem{DKS} 
J.R. Dorfman, T.R. Kirkpatrick, J.V. Sengers,
Annu. Rev. Phys. Chem. {\bf 45}, 213
(1994).

\bibitem{LL}L. D. Landau, E. M. Lifshitz, {\sl Fluid mechanics},
  Pergamon Press 1987.

\bibitem{S1}
H. Spohn, J. Phys. {\bf A 16}, 4275 (1983).

\bibitem{SC} 
R. Schmitz, Phys. Rep. {\bf 171}, 1 (1988).

\bibitem{K} J. Keizer {\sl Statistical thermodynamics of
  nonequilibrium processes}, Springer 1987.

\bibitem{CDK}
T.R. Kirkpatrick, E.G.D. Cohen, J.R. Dorfman,
Phys. Rev. A {\bf 26}, 950, 972, 995 (1982). 

\bibitem{BDGJL1}
L. Bertini, A. De Sole, D. Gabrielli, G. Jona-Lasinio, C. Landim,
Phys. Rev. Lett. {\bf 87}, 040601 (2001). 

\bibitem{BDGJL2} 
L. Bertini, A. De Sole, D. Gabrielli, G. Jona-Lasinio, C. Landim,   
J. Statist. Phys. {\bf 107}, 635
(2002).

\bibitem{E} 
A. Einstein, 
Annalen der Physik, {\bf 33}, 1275
(1910).
English translation in \emph{The collected papers of Albert Einstein}, 
vol.3 p. 231--249, Princeton University Press, 1993.

\bibitem{DLS2} 
B. Derrida, J.L. Lebowitz, E.R. Speer, 
J. Statist. Phys. {\bf 107}, 599
(2002).

\bibitem{BGL} 
L. Bertini, D. Gabrielli, J.L. Lebowitz, 
J. Statist. Phys. {\bf 121}, 843 (2005).

\bibitem{DE}
C. Enaud, B. Derrida, 
J. Statist. Phys. {\bf 114}, 537 (2004).

\bibitem{NOI9} 
L. Bertini, A. De Sole, D. Gabrielli, G. Jona-Lasinio, C. Landim, 
\emph{Stochastic interacting particle systems out of equilibrium}, 
Preprint arXiv:0705.1247 2007. 

\bibitem{DELO} 
B. Derrida, C. Enaud, C. Landim, S. Olla, 
J. Stat. Phys. {\bf 118}, 795 (2005).  

\bibitem{KMP} 
C. Kipnis, C. Marchioro, E. Presutti,
J. Statist.  Phys. {\bf 27}, 65 (1982).

\bibitem{BJ}
G. Basile, G. Jona-Lasinio,
Inter.\ J.\ Mod.\ Phys.\ B {\bf 18}, 479 (2004).


\end{thebibliography}
\end{document}